\documentclass[a4paper,11pt]{article}
\usepackage{jcappub} 
\usepackage{xcolor}
\usepackage{lineno}
\usepackage{subcaption}
\usepackage{amsmath}
\usepackage{bm}
\usepackage{float}
\usepackage{textcomp}

\newcommand{{\pasj}}{Publications of the Astronomical Society of Japan}

\title{ Impact of the Epoch of Reionization sources on the 21-cm bispectrum}
\arxivnumber{2406.03118}
\author[a]{Leon Noble,}
\author[b]{Mohd Kamran,}
\author[a,c]{Suman Majumdar,}
\author[a]{Chandra Shekhar Murmu,}
\author[d,e]{Raghunath Ghara,}
\author[f]{Garrelt Mellema,}
\author[g]{Ilian T. Iliev}
\author[c]{and Jonathan R. Pritchard}

\affiliation[a]{Department of Astronomy, Astrophysics \& Space Engineering, Indian Institute of Technology Indore, Indore 453552,  India}
\affiliation[b]{Department of Physics and Astronomy, Uppsala University, Uppsala 75237, Sweden}
\affiliation[c]{Department of Physics, Blackett Laboratory, Imperial College, London SW7 2AZ, UK}
\affiliation[d]{Haverford College, 370 Lancaster Ave, Haverford PA, 19041, USA}
\affiliation[e]{Center for Particle Cosmology, Department of Physics and Astronomy, University of Pennsylvania, Philadelphia, PA 19104, USA}
\affiliation[f]{The Oskar Klein Centre and The Department of Astronomy, Stockholm University, AlbaNova, SE-10691 Stockholm, Sweden}
\affiliation[g]{Department of Physics and Astronomy, Pevensey II Building, University of Sussex, Brighton
BN1 9QH, UK}

\emailAdd{leonnoblek@gmail.com}

\abstract{
The morphology of the 21-cm signal emitted by the neutral hydrogen present in the intergalactic medium (IGM) during the Epoch of Reionization (EoR) depends both on the properties of the sources of ionizing radiation and on the underlying physical processes within the IGM. Variation in the morphology of the IGM 21-cm signal due to the different sources of the EoR is expected to have a significant impact on the 21-cm bispectrum, which is one of the crucial observable statistics that can evaluate the non-Gaussianity present in the signal and which can be estimated from radio interferometric observations of the EoR. Here we present the 21-cm bispectrum for different reionization scenarios assuming different simulated models for the sources of reionization. We also demonstrate how well the 21-cm bispectrum can distinguish between different IGM 21-cm signal morphologies, arising due to the differences in the reionization scenarios, which will help us shed light on the nature of the sources of ionizing photons. 
Our estimated large-scale bispectrum for all unique $k$-triangle shapes shows a significant difference in the magnitude and sign across different reionization scenarios. Additionally, our focused analysis of bispectrum for a few specific $k$-triangle shapes (e.g. squeezed-limit, linear, and shapes in the vicinity of the squeezed-limit) shows that the large scale 21-cm bispectrum can distinguish between reionization scenarios that show inside-out, outside-in and a combination of inside-out and outside-in morphologies. These results highlight the potential of using the 21-cm bispectrum for constraining different reionization scenarios.
}
\begin{document}
\maketitle
\flushbottom
\section{Introduction}
\label{sec:intro}
Cosmic Dawn and  Epoch of Reionization (CD-EoR) remains one of the least well-known epochs in the evolutionary history of our universe. During this period, the radiation from the first luminous sources heated and almost completely ionized the neutral hydrogen (HI) present in the intergalactic medium (IGM). Over the past few decades, a considerable theoretical understanding of this epoch has developed, but these models need to be tested using observations. Our current understanding of this epoch is based on various indirect observations, including the measurement of Thomson scattering optical depth from Cosmic Microwave Background (CMB) observations~\cite{komastsu_2011,Planck_reionzation_history_2016,Planck_2018_cosmological_parameters}, absorption spectra of high redshift quasars~\cite{Becker_2001,Fan_2003,Barnett_2017}, and the luminosity function and clustering properties of \text{Lyman}-$\alpha$ emitters~\cite{Ouchi_2010,Zheng_2017,Taylor_2021}. However, these observations provide only a partial picture and fail to answer fundamental questions regarding the formation and evolution of first luminous sources, the properties of sources of ionizing radiation, the topology of the ionized regions during different stages of reionization and the detailed timeline of reionization. We need direct observation of the first luminous sources and the evolving IGM to answer these fundamental questions.
\par
The 21-cm radiation emitted by HI in the IGM due to the hyperfine spin-flip transition is an excellent probe to study CD-EoR~\cite{Pritchard_2012}. It is expected to carry a wealth of information about the properties of the sources of ionizing radiation, the state of the IGM, underlying physical processes within the IGM and the topology of ionized regions at different stages of reionization. A large number of ongoing radio interferometric experiments including GMRT~\cite{Paciga_2013}, LOFAR~\cite{Mertens_2020}, MWA~\cite{Barry_2019}, HERA~\cite{Hera_EoR_upper_limit_2022} and the upcoming SKA~\cite{Koopmans_2015} is sensitive enough to measure the fluctuations in the 21-cm signal via different summary statistics including the variance~\cite{Patil,Harker_2009,Watkinson_2015,Shimabukuro_2015,Kubota_2016}, power spectrum~\cite{Jensen_2013} etc. However, all of these experiments have only been able to achieve upper limits on the spherically averaged 21-cm power spectra at a range of redshifts of reionization and for a few specific length scales~\cite{Munshi_et_al_2024,Hera_impr_EoR_upper_limit_2022,Hera_EoR_upper_limit_2022,Trott_2020,Mertens_2020}. These upper limits on the power spectrum have allowed to rule out non-viable CD-EoR models \cite{Hera_EoR_upper_limit_2022,Ghara_2021,Greig_2021,Mondal_2020,Ghara_2020}.
\par
The power spectrum, by definition, measures the amplitude of the signal fluctuations at different scales and only completely characterizes the statistical properties of a signal if it is a Gaussian random field. However, the redshifted 21-cm radiation from the CD-EoR is highly non-Gaussian, and the level of non-Gaussianity varies with the nature of ionizing sources, state of the IGM and the physical processes within the IGM~~\cite{Bharadwaj_2005_bispectrum,Iliev_2006,Mellema_2006}. To capture the non-Gaussianity present in the 21-cm signal from the CD-EoR, one has to choose a statistic
sensitive to that non-Gaussianity. One-point statistics such as skewness and kurtosis~\cite{Harker_2009,Watkinson_2014,Watkinson_2015,Shimabukuro_2015,Kubota_2016} are one such possibility. However, these one-point statistics can only capture the non-Gaussianity at a single length scale and cannot quantify the correlation that might be present in the signal fluctuations at different scales. To achieve the latter, one has to consider higher-order statistics such as the bispectrum~\cite{Majumdar2018,Hutter_2020,Majumdar_2020,kamran_2020,Rajeshbispectrumdetectability,Watkinson_2019,Ma_2021,KamranPRL2021,Kamran_2022,Raste_2023,Shimabukuro_2016,Watkinson_2022,Tiwari_2022}. 
\par
In a previous article~\cite{Majumdar2018}, using a set of simulated 21-cm signals from the EoR, we showed that this signal is highly non-Gaussian due to the growth of the ionized region in the neutral IGM and this non-Gaussianity evolves with time as the reionization progresses. We identified that at the earlier stages of the reionization, the sign of the EoR 21-cm bispectrum depends on whether the non-Gaussianity in the 21-cm signal is dominated by the non-Gaussianity of the matter density field or by the HI field. The EoR 21-cm bispectrum for all unique $k$-triangle shapes was first explored in \cite{Majumdar_2020}. They showed that the sign and magnitude of the 21-cm bispectrum depend on the size and shape of the $k$-triangle as well as the stage of reionization. The line-of-sight anisotropies, such as Redshift Space Distortions (RSD) and the Light Cone (LC) effect, impact the sign and magnitude of the EoR 21-cm bispectrum \cite{Majumdar_2020, Gill_2023,Rajeshbispectrumdetectability}. In~\cite{Hutter_2020}, the authors studied the impact of different inside-out reionization models on the EoR 21-cm bispectrum, and they identified that the 21-cm bispectrum is sensitive to the variation in the reionization topology. Recent works \cite{Gill_2023} and \cite{Raste_2023} reveal that the 21-cm bispectrum is able to capture important topological transitions during reionization. The sign change in the 21-cm bispectrum at a later stage of the reionization indicates the stage during which the neutral HI islands start to appear in the ionized background. However, \cite{Gill_2023} restricted their analysis to a single inside-out reionization model, and \cite{Raste_2023} focused on only two different reionization histories. 
\par
Several previous works demonstrated that the change in the properties of the sources of ionizing radiation and physical processes within the IGM affects the morphology of the IGM 21-cm signal from the EoR \cite{Majumdar_2016,Mesinger_2013,Watkinson_2014,Sobacchi_2014,Geil_2016,Kulkarni_2017,Hassan_2018,Eide_2020,Pathak_2022, Cain_2023,Lu_2024}. Thus the non-Gaussianity in the signal and its time evolution similarly vary depending on the reionization scenario. Earlier studies of the EoR 21-cm bispectrum~\cite{Majumdar2018,Majumdar_2020,Shimabukuro_2016,Raste_2023,Gill_2023} did not explore the impact of varying properties of the sources of ionizing radiation and the variations in the physical processes within the IGM on this signal statistic. Here we study the impact of varying source properties on the 21-cm bispectrum and we ask the question, to what extent the 21-cm bispectrum can distinguish between different reionization scenarios?
We address this point based on a set of simulated 21-cm maps representing different reionization scenarios, which mimic the inside-out, outside-in, and combination of inside-out and outside-in reionization morphologies. Depending on the morphology of the IGM 21-cm signal, the sign and magnitude of the EoR 21-cm bispectrum will vary. The morphology of 21-cm maps, in turn, is connected to the nature of ionizing sources and IGM properties. The different source models considered in this work differ by how the number of ionizing photons is related to the host halo mass and the distribution of the rest frame energy of these photons. We analyze the 21-cm bispectrum of all unique $k$-\text{triangle} shapes for all of these simulated reionization scenarios. We finally explore how well one can distinguish between different reionization scenarios using the size, shape, sign and magnitude of the 21-cm bispectrum. 
\par
The article is organized as follows: in section~\ref{section: reionization simulations}, we discuss the simulations used to create different source models and various reionization scenarios. Section \ref{section:  bispectrum estimation} describes the formalism used for estimating the bispectrum from simulated 21-cm maps. In section \ref{sec:results}, we present our bispectrum results for all of the reionization scenarios. Finally, in section~\ref{section: summary}, we summarise our results. Throughout this paper, we used WMAP ﬁve year data release cosmological parameters $h=0.7$, $\Omega_m=0.27$, $\Omega_{\Lambda}=0.73$, $\Omega_b h^2=0.0226$~\cite{Komatsu_2009}.

\section{Reionization simulations}
\label{section: reionization simulations}
In order to capture the complexity of reionization, which is affected by the astrophysics at small length scales as well as the matter distribution at large length scales, one has to simulate it in large enough volumes (at least $\approx (\text{200 Mpc})^{3})$~\cite{Iliev_2014}. Simulating reionization in large cosmological volumes is further motivated by the fact that several ongoing (e.g. LOFAR) and upcoming (e.g. SKA-Low) radio interferometric experiments will have fields of views that translate into observing volumes of the order of $\geq (0.5\, {\rm Gpc})^3$ or more  \cite{Koopmans_2015}. A 3D numerical radiative transfer simulation can provide the most realistic maps of 21-cm signal from reionization. These simulations incorporate detailed physical processes happening at the sources and in the IGM in a self-consistent fashion. However, these simulations become computationally expensive because of the incorporation of these multi-scale (with large dynamic range) physical processes. Thus, to simulate the signal at large cosmological volumes and/or to re-simulate the signal for different sets of reionization parameters demands a huge amount of computing resources and time. On the other hand, computationally less expensive semi-numerical methods based on excursion set formalism~\cite{Furlanetto_2004} can simulate the 21-cm signal from the reionization with approximate physics at a much lower computational cost. This allows us to re-simulate the signal for a range of reionization scenarios. For our analysis in this article, we use a set of semi-numerically simulated 21-cm signals from \cite{Majumdar_2016}. Here we very briefly describe the relevant details of these simulations and source models and refer the reader to \cite{Majumdar_2016} for further detail. 
\par
The simulated 21-cm signal presented in \cite{Majumdar_2016} is based on a CUBEP$^{3}$M N-body ~\cite{Harnois_2013} run, which was used to generate the input underlying dark matter distribution and halo catalogues at a series of redshifts. The size of the simulation cube along each side is $500\, h^{-1}=714.28$ cMpc (comoving Mpc). These simulations had $6912^{3}$ dark mattter particles, each of mass $4.0\times10^{7} M_\odot$ on a $13824^{3}$ mesh. The dark matter field was then down-sampled to a $600^{3}$ grid for simulating reionization. The 21-cm signal from reionization is thus generated with a grid spacing of  $1.19$ cMpc. The halo catalogues generated from these dark matter fields have a minimum halo mass of $2.02\times10^{9}M_{\odot}$. 
The dark matter field and the halo catalogue are then used as inputs to a semi-numerical simulation \cite{Majumdar_2014} to simulate the resulting 21-cm fields for various reionization scenarios presented in \cite{Majumdar_2016}. All the length scales mentioned in this article are in comoving units, so hereafter we drop the cMpc notation in favor of Mpc.

\subsection{Reionization source models and scenarios}
\label{subsection:source models and scenarios}
Here we briefly describe the reionization source models and scenarios presented in \cite{Majumdar_2016} and used for 21-cm bispectrum analysis in this article. We focus on creating diverse IGM 21-cm maps showing significant variations in the 21-cm morphology to study their impact on the target signal statistics, i.e., the 21-cm bispectrum, and it is not to create more realistic 21-cm maps based on the various observational constraints on the source population.

\subsubsection{Source models}
\label{sec:source}
\begin{itemize}
  \item Ultraviolet photons from the galaxies~(UV): In this source model, we consider the far ultraviolet stellar radiation from the galaxies residing in the dark matter halos as the source of the ionizing photons. We assume that the total number of ionizing photons~$N_{\gamma}$ emitted into the intergalactic medium from each galaxy by the time considered is directly proportional to the mass of the halo~$M_{h}$ that hosts that galaxy~\cite{Majumdar_2014,Choudhury_2009}
\begin{align}
    \label{Equation_1}
    N_{\gamma}=N_{\rm ion}\frac{M_{h}\Omega_b}{m_{p}\Omega_m}.
\end{align}
$N_{\rm ion}$ is a dimensionless constant representing the number of ionizing photons produced in the stars and entering the IGM per baryon inside the halo, and $m_{p}$ is the mass of the proton. The $N_{\rm ion}$ depends on many factors, such as the fraction of baryons in the stars and the fraction of ionizing photons produced by stars that escape into the IGM etc. Tuning of $N_{\rm ion}$ effectively can change the speed of reionization. This particular case where $N_{\gamma} \propto M_{h}$ is labeled as the fiducial reionization scenario used in this study.

 \item Uniform ionizing background~(UIB): This model is different from the fiducial scenario, such that the ionizing photons are hard X-rays assumed to be emitted by sources like AGNs, high-mass X-ray binaries etc. These X-ray photons can easily escape the host galaxy and have a long mean free path. Due to the large mean free path of hard X-rays, it contributes to a uniform ionizing background, resulting in a reionization morphology where low density regions ionize first and high density regions ionize later. 

 \item Soft X-ray photons~(SXR): In this model, the ionizing photons are soft X-rays distributed uniformly in a region around the source, where the total number of ionizing photons produced follows a similar relation as equation \eqref{Equation_1}. The region is limited by the mean free path of the soft X-rays determined by the prescription of \cite{Mesinger_2013}, resulting in a uniform ionizing background around the source.

 \item Power law mass-dependent efficiency~(PL): In this model, we assume that the total number of ionizing photons in the ultraviolet part of the spectrum entering the IGM from all the halos  follows the relationship
\begin{align}
    N_{\gamma} \propto M_{h}^{n}.
\end{align}
Here, we considered two scenarios where the power law index is  $n=2$ and $n=3$. 
\end{itemize}
\begin{figure}
    \centering    \includegraphics[width=\textwidth]{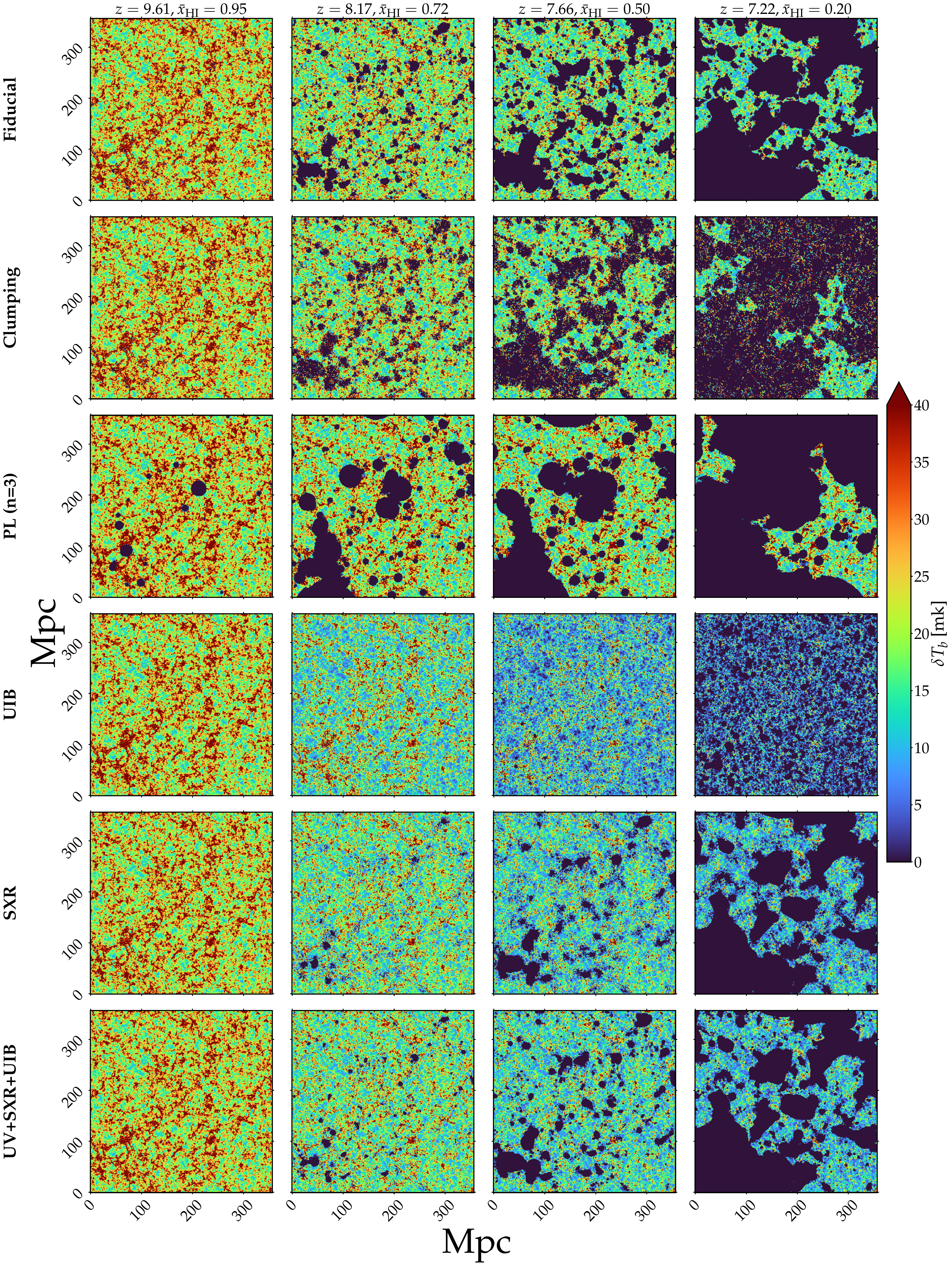}
    \caption{Zoomed-in slices (side length 357 Mpc) of simulated 21-cm maps for six different reionization scenarios taken from \cite{Majumdar_2016} is shown in this figure. Fiducial, clumping, and PL($n=3$) are “inside-out” reionization scenarios, UIB is “outside-in” reionization scenario, and SXR and UV+SXR+UIB are reionization scenarios that show a combination of inside-out and outside-in morphology.}
    \label{21-cm_map}
\end{figure}
\subsubsection{Reionization scenarios}
In this study, we use the 21-cm maps of seven different reionization scenarios from~\cite{Majumdar_2016}. These scenarios are various combinations of different source models described in Sec \ref{sec:source}. The same set of 21-cm maps was also used in \cite{Pathak_2022}, where they studied how different reionization scenarios can be distinguished using the Largest Cluster Statistics, an image-based statistic. The fiducial scenario is the one with UV photons contributing $100\%$ of ionizing photons. In this scenario, the high-density HI regions around the sources are ionized first, and low-density regions are ionized at a later stage of the reionization. This type of reionization morphology is typically known as inside-out reionization \cite{Iliev_2006,Choudhury_2009}. All of the reionization scenarios except the clumping scenario considered in this work to assume a uniform rate of recombination of HI in the IGM. In the clumping scenario, density-dependent recombination is included following the prescription of \cite{Choudhury_2009}.  
Even though the $100\%$  ionizing photons are UV photons in the PL ($n=2$) and PL ($n=3$) scenarios, the morphology of the 21-cm maps varies significantly from the fiducial scenario. In these scenarios, high-mass halos contribute more than low-mass halos in ionizing the IGM. In the UIB and SXR scenarios, $80\%$ of ionizing photons are hard X-rays and soft X-rays respectively, and the remaining photons belong to the UV part of the spectrum. The ionizing background produced by the UIB scenario is uniform due to the large mean free path of the hard X-rays. In this scenario, the reionization progresses outside-in \cite{Choudhury_2009}, where the low-density HI regions get ionized first and the high-density HI regions get ionized at a later stage of the reionization. The morphology of the IGM 21-cm signal of SXR-dominated and UV+SXR+UIB ($50\%$ UV photons, $40\%$ soft X-rays, and $10\%$ hard X-rays) scenarios are in between inside-out and outside-in. Finally, for all of these scenarios, the parameter $N_{\rm ion}$ is tuned so that all of them follow the same reionization history. This is to ensure that any change in the resulting 21-cm morphology between different reionization scenarios is only due to the differences in source models and not due to the differences in the underlying matter distribution. 
\section{Bispectrum estimation}
\label{section:  bispectrum estimation}
\subsection{Estimation of bispectrum from simulated 21-cm data}
The bispectrum $B(\mathbf{k_1},\mathbf{k_2},\mathbf{k_3})$ of the 21-cm  differential brightness temperature field $\delta T_{b}(\mathbf{x})$ is defined as 
\begin{align}
    \langle \Delta_{b}(\mathbf{k_1}) \Delta_{b}(\mathbf{k_2}) \Delta_{b}(\mathbf{k_3}) \rangle =V \delta_{\mathbf{k_1} + \mathbf{k_2} + \mathbf{k_3},0}~B(\mathbf{k_1},\mathbf{k_2},\mathbf{k_3}),
\end{align}
where the $\Delta_{b}(\mathbf{k})$ is the Fourier transform of the differential brightness temperature field $\delta T_{b}(\mathbf{x})$ and V is the total volume of the simulation box. $\delta_{\mathbf{k_1} + \mathbf{k_2} + \mathbf{k_3},0}$ is the Kronecker delta function, and its numerical value is equal to one when the condition  $\mathbf{k_1} + \mathbf{k_2} + \mathbf{k_3}=0$ satisfied and zero otherwise. This ensures that only the closed $k$-triangles contribute to the bispectrum. The bispectrum depends on the triangle shape and the size of the triangles in the Fourier space. From the simulated data cube, we estimate the binned bispectrum by following the algorithm from~\cite{Majumdar2018,Majumdar_2020}. The binned bispectrum $\bar B_{i}(\mathbf{k_1},\mathbf{k_2},\mathbf{k_3})$ of the i$^{\text{th}}$ triangle configuration is given by 
\begin{align}
\bar B_{i}(\mathbf{k_1},\mathbf{k_2},\mathbf{k_3})=\frac{1}{VN_{\rm tri}} \sum_{[\mathbf{k_1}+\mathbf{k_2}+\mathbf{k_3}=0]~\epsilon~i}    \langle \Delta_{b}(\mathbf{k_1}) \Delta_{b}(\mathbf{k_2}) \Delta_{b}(\mathbf{k_3}) \rangle,
\end{align}
where $N_{\rm tri}$ is the total number of statistically independent closed $k$-triangles that belongs to the i$^{\text{th}}$ bin. For computational ease, the bispectrum algorithm also follows two additional constraints, which define the  shape of the $k$-triangles  
\begin{gather}
    t=\frac{k_2}{k_1},\\
    \cos\theta=-\frac{\mathbf{k_1}\cdot\mathbf{k_2}}{k_1  k_2},
\end{gather}
where $k_1$ and $k_2$ are the magnitude of the vectors $\mathbf{k_1}$ and $\mathbf{k_2}$ respectively and $\theta$ is the angle between the vectors $\mathbf{k_1}$ and $\mathbf{k_2}$ as shown in the left panel of  Figure~1 in~\cite{Majumdar_2020}.
\subsection{The unique triangle configurations in the triangle parameter space}
For $k$-triangles  of a specific size, determined by the magnitude of the $k$ mode, the shape of the $k$-triangles in the $t-\cos\theta$ space can be uniquely determined by following the formalism introduced in ~\cite{Bharadwaj_2020}, i.e.,
\begin{gather}
    k_{1}\geq k_{2}\geq k_{3},\\
    0.5\leq t\leq 1.0,\\
    0.5\leq \cos\theta\leq 1.0~.
\end{gather}
The $k$-triangles that satisfy these conditions are confined to the region where $t-\cos\theta \geq 0.5$ in the $t-\cos\theta$ parameter space as shown in the right panel of Figure~1 in \cite{Majumdar_2020}. In this article, we will mostly explore the ability of linear ($ 0.5\leq t\leq1.0$ and $\cos\theta=1.0$) and L-isosceles ( $t=1$ and $ 0.5\leq \cos\theta\leq1.0$~) $k$-triangle bispectra in distinguishing the reionization models.  We discuss the motivation behind this choice of $k$-triangles in Section \ref{sec:results}. 
\par
We estimated the spherically averaged bispectrum on the simulated data cube for different redshifts corresponding to reionization. We divided the entire $t-\cos\theta$ parameter space with grid size $\Delta t=0.05$ and $\Delta \cos\theta=0.01$. The $k_1$-range is binned into 15 logarithmic bin with $k_1$ varying from $k_{\min}=2\pi/\text{box size}$ to $k_{\rm max}=2\pi/\text{grid spacing}$. Additionally, to reduce the sample variance, we averaged the bispectrum over five $\cos\theta$ bins, which resulted in the final grid resolution of $\Delta\cos\theta=0.05$.
\section{Results}
\label{sec:results}
We next discuss the 21-cm bispectrum for the seven different reionization scenarios considered in this work. The power spectrum, by definition, cannot capture the non-Gaussianity present in the signal. Hence, we have to rely on higher-order statistics such as bispectrum which can quantify the non-Gaussianity present in the signal. Therefore, the natural question to ask is whether the additional information captured in the bispectrum will help to outperform the power spectrum to distinguish different reionization scenarios. Thus, before we start the discussion on the 21-cm bispectrum, we first review the nature of the 21-cm power spectrum for different reionization scenarios (see section~\ref{subsection:21-cm_power_spectrum}). While discussing the 21-cm bispectrum results (section \ref{subsection:21-cm_bispectrum_fiducial} onwards), we mainly focus on the 21-cm bispectrum of a few specific unique $k$-triangle shapes such as the squeezed-limit, equilateral, stretched, L-isosceles and linear. Additionally, we mainly focus on the 21-cm bispectrum at large scales ($0.16~\text{Mpc}^{-1}\leq k_1\leq0.33~\text{Mpc}^{-1}$ ). The choice of scales and specific shapes of $k$-triangle is motivated by the earlier studies, which have shown that the 21-cm bispectrum estimated at large scales (low $k_{1}$ values) has the maximum detectability via future SKA1-Low surveys ~\cite{Rajeshbispectrumdetectability}. Among those $k$-triangles, the squeezed-limit has the highest signal-to-noise ratio followed by the linear $k$-triangles for large scales and the squeezed-limit has the highest magnitude among all the unique $k$-triangles at all stages of reionization~\cite{Majumdar_2020}. We notice that most of the unique $k$-triangles in the vicinity of the linear and L-isosceles show more features in the 21-cm bispectrum than other $k$-triangles. In section~\ref{subsection:21-cm_bispectrum_fiducial}, we discuss the 21-cm bispectrum for the fiducial scenario, followed by the interpretation of the nature of the bispectrum in section~\ref{subsection:interpreting_21-cm_bispecctrum_fiducial}. In section~\ref{sec:bispec_reion_scenario} and \ref{sec:bispec_difference}, we quantify the differences in the 21-cm bispectrum between different reionization scenarios with respect to the fiducial case.
\subsection{21-cm power spectrum}
\label{subsection:21-cm_power_spectrum}
\begin{figure}
    \centering
    \includegraphics[width=\textwidth]{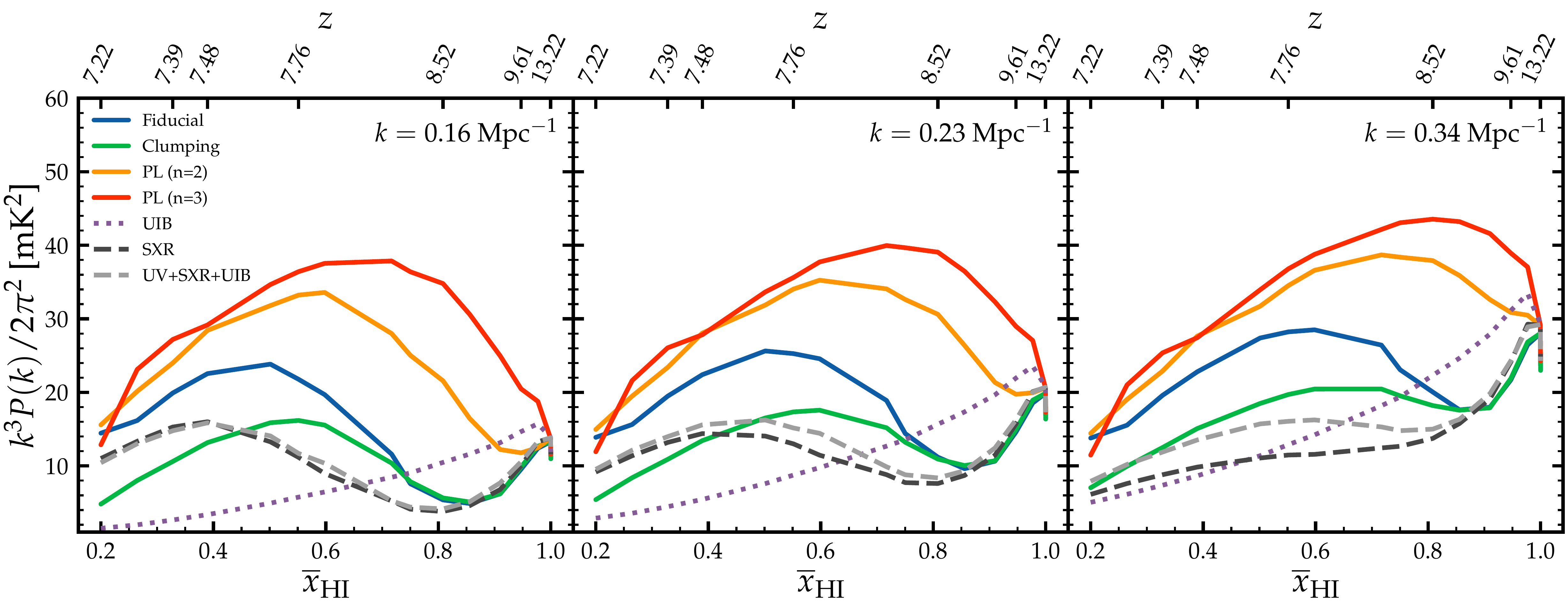}
     \caption{Evolution of the 21-cm power spectrum for different reionization scenarios with global neutral fraction (redshift) is shown in this figure. Solid, dotted, and dashed lines represent reionization scenarios that show inside-out, outside-in, and a combination of inside-out and outside-in morphologies respectively.}
    \label{power_spectrum}
\end{figure}

In this section, we review the impact of different reionization scenarios on the 21-cm power spectrum. A thorough discussion of these 21-cm power spectra can be found in \cite{Majumdar_2016}. Figure \ref{power_spectrum} shows the evolution of the 21-cm power spectrum with the evolving state of ionization of the IGM for $k=0.16~\text{Mpc}^{-1}$. First, we discuss the nature of the 21-cm power spectrum for the fiducial reionization scenario. Considering the fiducial scenario as our benchmark, we compare the signal power spectra for other reionization scenarios.  The 21-cm brightness temperature fluctuations are initially driven by variations in the HI distribution, which follows the matter density field. As ionized bubbles begin to form, the magnitude of the 21-cm power spectrum starts decreasing and reaches a minimum around neutral fraction $\bar x_{\rm HI}\approx 0.85$. This is due to the fact that in the inside-out reionization scenario, the high-density HI region around the halo gets ionized first, which leads to a loss of contrast in brightness temperatures between low and high-density regions, which results in a drop in the 21-cm power spectrum, as pointed out in \cite{Lidz_2008, Ivelin_Georgiev_2022}. During the intermediate stages of the reionization, the ionized bubbles grow in size and merge to form large ionized regions. 
This results in increased fluctuations in the 21-cm signal, which in turn results in the rise of the 21-cm power spectrum and leads to a peak in the power spectrum at around the time when half of the volume is ionized. At this stage, the 21-cm power spectrum no longer follows the matter power spectrum but is rather driven by the contribution from the neutral fraction field. Throughout the later stages of reionization, the neutral fraction field remains the dominant contributor to the 21-cm power spectrum. During the late stages of the reionization, the 21-cm power spectrum starts to decline as the volume of the interconnected single large ionized region increases abruptly and fills in most of the IGM. The power spectrum for $k=0.23\, {\rm and}\, 0.34\,~\text{Mpc}^{-1}$ (central and right panels of Figure \ref{power_spectrum}) qualitatively show an almost similar time evolution as that of the $k=0.16~\text{Mpc}^{-1}$. The only difference at these scales is the change in the overall amplitude of the power spectrum and a small shift in time for various distinguishing features in the power spectrum.
\par
Next, we compare the nature of the 21-cm power spectrum for various other reionization scenarios (represented by different line styles in Figure \ref{power_spectrum}) considered in this work with that of the fiducial scenario. During the early stages of the reionization, the 21-cm power spectrum for the clumping scenario follows the power spectrum for the fiducial scenario. However, the 21-cm power spectrum of PL scenarios show different trends during the early stages of reionization and have a larger magnitude of 21-cm power than that of the fiducial scenario. The dip in the 21-cm power spectrum in the early stage of reionization is absent in the PL  scenarios, as the contribution from fluctuations in the neutral fraction field to the power spectrum starts to dominate at a very early stage of the reionization due to the larger ionized regions forming at the very beginning of reionization. During the intermediate stages of the reionization, the magnitude of the 21-cm power spectrum for  PL scenarios is much higher than that of the fiducial scenario, as the typical size of the ionized bubbles is large in the former case. In these scenarios, the peak in the power spectrum appears at a higher neutral fraction. This is due to the fact that more than half of the volume of the IGM is fully ionized at a lower mass-averaged neutral fraction.
\par
The reionization scenarios which lie in between the extremes of inside-out and outside-in, such as SXR and UV+ SXR+UIB scenarios, show similar features in the 21-cm power spectrum as shown in the fiducial case. However, these features are shifted towards the later stage of reionization for these cases compared to the fiducial one. The peak in the power spectrum shifts towards a lower neutral fraction value because more than half of the volume of the IGM is fully ionized at a lower neutral fraction. These scenarios have a lower 21-cm power spectrum at all stages of reionization due to the presence of a large volume of partially ionized region. Among all reionization scenarios considered here, the UIB scenario shows an entirely different nature of power spectrum evolution. At all stages of reionization, the UIB scenario power spectrum closely follows the matter density power spectrum (with a modulation in amplitude). This is because the contribution from the neutral fraction field to the 21-cm power spectrum is negligible compared to the matter density field due to the presence of a uniform ionizing background in the UIB scenario.

\subsection{21-cm bispectrum for the fiducial scenario}
\label{subsection:21-cm_bispectrum_fiducial}
\begin{figure}
    \centering
    \includegraphics[width=\textwidth]{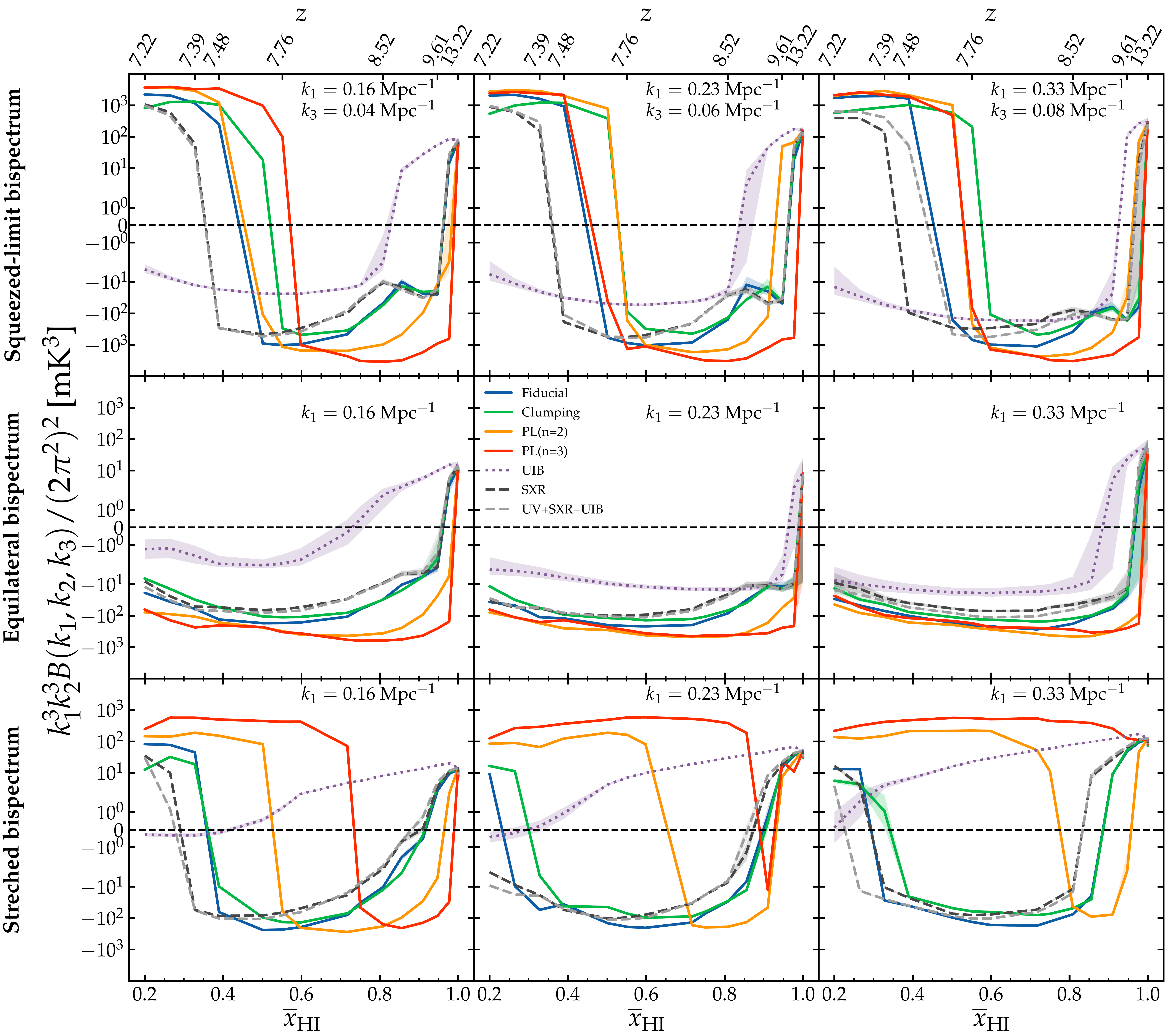}
    \caption{Evolution of the 21-cm  bispectrum for squeezed-limit (top panel), equilateral (middle panel), and stretched (bottom panel) $k$-triangles for different reionization scenarios with global neutral fraction (redshift) is shown in this figure. Solid, dotted, and dashed lines represent reionization scenarios that show inside-out, outside-in, and a combination of inside-out and outside-in morphologies respectively. The shaded region around the line plot shows the $3\sigma$ error in
    the bispectrum due to instrumental noise and sample variance for 1000 hours of observation with SKA1-Low.}
    \label{SQ_BS}
\end{figure}

We first discuss the 21-cm bispectrum of the fiducial scenario. The top, middle, and bottom panels of Figure \ref{SQ_BS} show the bispectra for squeezed-limit, equilateral, and stretched $k$-triangles, respectively. First we discuss the 21-cm bispectrum for squeezed-limit $k$-triangles for $k_{1}=0.16~\text{Mpc}^{-1}$. The solid blue line in the left top panel of Figure \ref{SQ_BS} shows the evolution of this bispectrum with neutral fraction (redshift). Before reionization begins, the sign of the 21-cm bispectrum for the squeezed-limit $k$-triangles is positive. During the early stages of reionization, when isolated ionized regions begin forming around halos, we observe a decrease in the magnitude of the bispectrum, and later, the bispectrum shows a sign change from positive to negative around neutral fraction $\bar x_{\rm HI }\approx 0.95$.
During the intermediate stages of reionization, when the ionized bubbles grow in size and isolated bubbles begin to coalesce to form large ionized regions, the bispectrum magnitude starts to increase. However, around global neutral fraction $\bar x_{\rm HI}\approx0.90$, we notice a reduction in the magnitude of the bispectrum similar to what we observed in the  21-cm power spectrum. The magnitude of the 21-cm bispectrum reaches a maximum around $\bar x_{\rm HI}\approx0.55$ and then starts decreasing as reionization progresses, during which the volume of ionized regions increases rapidly. Later on, around $\bar x_{\rm HI}\approx0.45$ bispectrum shows another sign change from negative to positive. This sign change appears when more than half of the volume of the IGM is fully ionized. During the late stages of the reionization, after which the bispectrum shows the second sign change, the magnitude of the bispectrum increases and becomes comparable to the magnitude of the bispectrum during the intermediate stages of the reionization. These features of the bispectrum are robust for $k_{1}$ range $0.16~\text{Mpc}^{-1} \leq  k_{1} \leq 0.33~\text{Mpc}^{-1}$. Further, as we vary the $k_1$ value from $k_{1}=0.16~\text{Mpc}^{-1}$ to $k_{1}=0.33~\text{Mpc}^{-1}$ the magnitude of the bispectrum increases during the intermediate stages of the reionization. 
\par
The middle left panel of Figure \ref{SQ_BS} shows the 21-cm bispectrum for equilateral $k$-triangles (solid blue line) for $k_{1}=0.16~\text{Mpc}^{-1}$. The bispectrum for equilateral $k$-triangles has fewer features than the same for squeezed-limit $k$-triangles. Similar to the features observed in squeezed-limit bispectrum during the early stages of reionization, the magnitude of the equilateral $k$-triangles bispectrum also decreases and eventually shows a sign change from positive to negative. The magnitude of the bispectrum for equilateral $k$-triangles increases during the intermediate stages of the reionization and during the late stages it starts to decrease. However, the bispectrum for equilateral $k$-triangles does not show a sign change at the late stages of reionization. 
We also study the 21-cm bispectrum for stretched $k$-triangles for $k_1=\{0.16,0.23,0.33\}~\text{Mpc}^{-1}$. At the early stages of the reionization, the bispectrum for the stretched $k$-triangles shows a similar trend following that of the squeezed-limit and equilateral $k$-triangles. However, the first sign change in the bispectrum is shifted towards a lower neutral fraction. Similar to the bispectrum for squeezed-limit $k$-triangles, the stretched $k$-triangles bispectrum also shows a sign change at a late stage of reionization. This sign change occurs at a neutral fraction lower than compared to that of the squeezed-limit $k$-triangles. Further, the bispectrum for stretched and equilateral $k$-triangles has a lower magnitude than squeezed-limit $k$-triangles bispectrum.
\begin{figure}
    \begin{subfigure}{0.5\textwidth}
        \includegraphics[width=\linewidth]{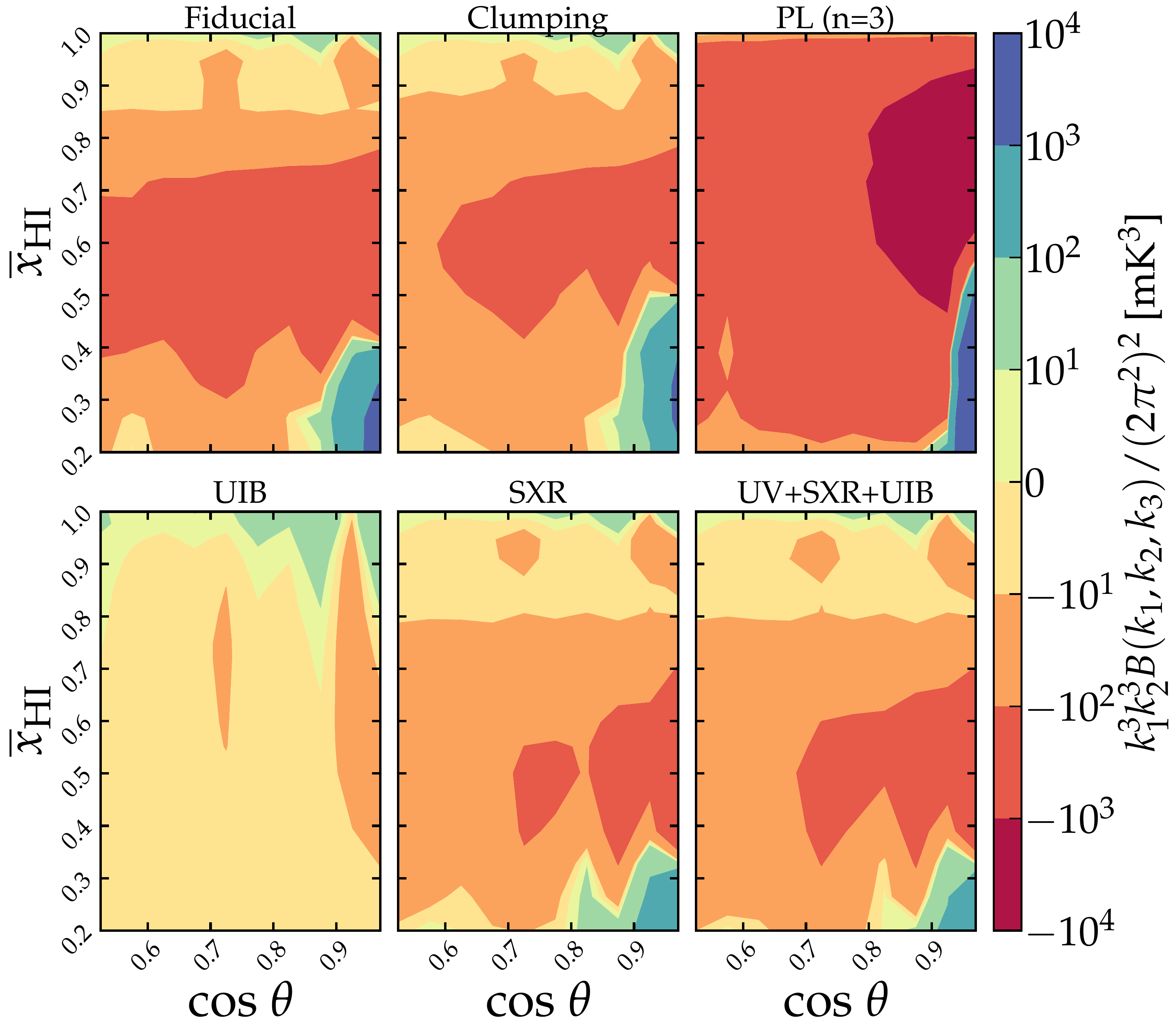}
    \end{subfigure}
    \hfill
    \begin{subfigure}{0.5\textwidth}
        \includegraphics[width=\linewidth]{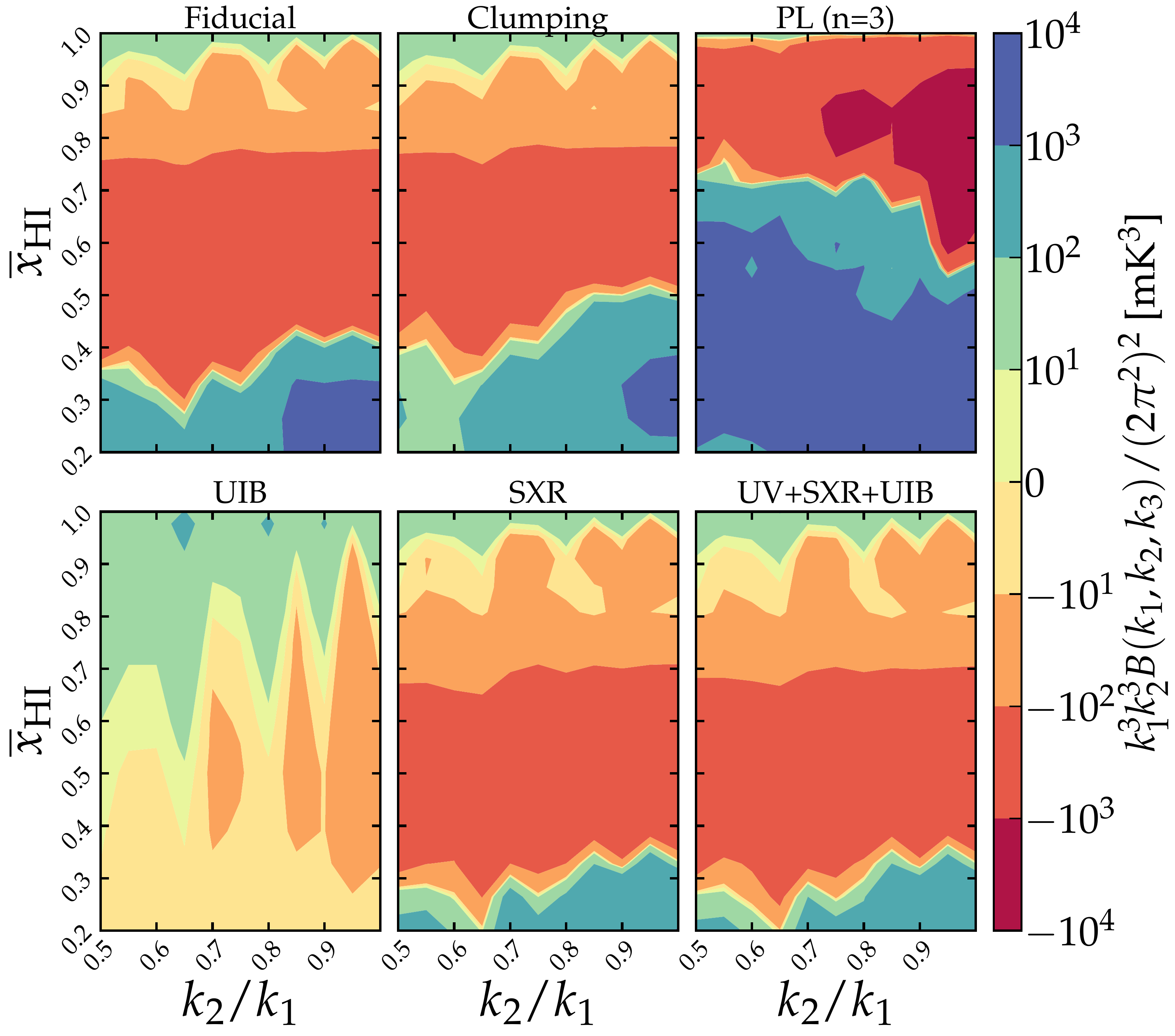}
    \end{subfigure}
    \hfill
    \caption{Left and right panel of this figure shows the 21-cm bispectrum for L-isosceles and linear $k$-triangles ($k_1=0.16~\text{Mpc}^{-1}$) respectively for six different reionization scenarios.}
    \label{L-isosceles_and_linear }
\end{figure}
\par
Next, we discuss the 21-cm bispectrum for the L-isosceles and linear $k$-triangles for $k_{1}=0.16~\text{Mpc}^{-1}$. First, we discuss the nature of the bispectrum for L-isosceles $k$-triangles of the fiducial scenario, which is shown in the left panel of Figure~\ref{L-isosceles_and_linear }. Recall that the squeezed-limit and equilateral $k$-triangles are special cases of the L-isosceles $k$-triangles. During the early and mid stages of the reionization, the bispectrum for the L-isosceles $k$-triangles follows the nature of squeezed-limit and equilateral $k$-triangles bispectra. However, at the late stages of the reionization, when half of the volume is ionized, the bispectrum for only L-isosceles $k$-triangles in the vicinity of the squeezed-limit $k$-triangles ($\cos \theta\in [0.8,1]$) shows a change in sign from negative to positive. This change in the sign starts appearing around $\bar x_{\rm HI}\approx 0.4$, and the magnitude of the bispectrum is highest at squeezed-limit $k$-triangles. This indicates that the bispectra for the $k$-triangles close to the squeezed-limit and the equilateral $k$-triangles are probing different features in the 21-cm signal fluctuations.
\par
The right panel of Figure~\ref{L-isosceles_and_linear } shows the bispectrum for the linear $k$-triangles. The bispectrum for the linear $k$-triangles closely follows the squeezed-limit $k$-triangles at all stages of the reionization. During the early stages of the reionization, the bispectrum of the linear $k$-triangles starts decreasing in magnitude and eventually shows the first sign change from positive to negative. Its magnitude increases during the intermediate stage of the reionization. During the late stages of the EoR, the sign of the bispectrum changes from negative to positive and its magnitude starts increasing. At this stage, the magnitude of the bispectrum is highest in the vicinity of the squeezed-limit $k$-triangles. All these bispectrum features observed for the fiducial scenario are in agreement with the previous independent studies \cite{Majumdar2018,Majumdar_2020,Gill_2023} which explored the EoR 21-cm bispectrum for reionization models similar to the fiducial scenario considered in this work. 
\subsection{Interpreting the 21-cm bispectrum for the fiducial scenario}
\label{subsection:interpreting_21-cm_bispecctrum_fiducial}
We next discuss our interpretation of the sign and sign change of the 21-cm bispectrum for the fiducial scenario. Earlier work on the CD 21-cm bispectrum showed that the sign of the large scale 21-cm bispectrum for the squeezed-limit $k$-triangle is determined by the relative contrast between the 21-cm signal fluctuations and the background~\cite{KamranPRL2021,Kamran_2022}.
To understand how the large-scale 21-cm bispectrum for the squeezed-limit $k$-triangle is able to capture the relative contrast between the 21-cm signal fluctuations and the background, we can use a simple toy model which was used in several previous works for the interpretation of the CD-EoR 21-cm bispectrum~\cite{BharadwajandPandey2005,Majumdar2018,KamranPRL2021,Kamran_2022,Gill_2023}.
\par
 At the earlier stages of reionization, more volume in the IGM is neutral than ionized. At this stage,  our toy model assumes that there are non-overlapping ionized regions with zero 21-cm signal in a uniform-density background of neutral hydrogen. Previously, \cite{BharadwajandPandey2005,Majumdar2018} showed that in this case, the Fourier transform of the signal fluctuations ($\Delta_{b}(\mathbf{k})$) will be negative. Thus, the sign of the squeezed-limit bispectrum (bispectrum being the product of the three $\Delta_{b}(\mathbf{k})$s) will also be negative. Note that this model is only valid at length scales larger than the typical size of the ionized region~\cite{BharadwajandAli2004,BharadwajandAli2005,Zaldarriaga_2004}. This explains the first sign change from positive to negative in the 21-cm bispectrum for squeezed-limit $k$-triangles for the fiducial scenario. This sign change can be interpreted as the tracer of the beginning of the reionization and marks the first important transition in the morphology of the IGM 21-cm signal. Here the large scale 21-cm bispectrum for the squeezed limit $k$-triangles captures the relative contrast between the signal fluctuations and the background. The IGM with 21-cm signal in emission acts as the background, and in this background, there are ionized regions with no signal.
 \par
The toy model discussed here remains valid even when one considers a mean subtracted 21-cm field as the magnitude and the sign of $\Delta_b(\mathbf{k})$ is determined by the contrast between the signal fluctuations and background, as discussed in \cite{KamranPRL2021,Kamran_2022}. 
Further note that we use this toy model to only predict the sign of the bispectrum, and we do not use it to determine the nature of the magnitude of the bispectrum at different stages of the reionization.
\par
We observe that the sign of the 21-cm bispectrum is positive during the late stages of the reionization. When half of the volume of the IGM is ionized, the sign of the bispectrum changes from
negative to positive (see Figure \ref{SQ_BS}).
At this stage of the reionization, the connected large ionized regions become the background. In this ionized background, there exist neutral regions with 21-cm signal in emission. In analogy with the first sign change of the bispectrum, the sign of the fluctuations at this stage will be positive, which yields a positive bispectrum.
\par
Next, we discuss the redshift evolution of the magnitude of the 21-cm bispectrum for squeezed-limit $k$-triangles for this scenario. 
During the early stages of reionization, when isolated ionized regions begin forming around halos, we observe a decrease in the magnitude of the 21-cm bispectrum. This trend can be explained as follows. The 21-cm bispectrum can be decomposed into the sum of neutral fraction bispectrum, matter density bispectrum and the sum of all the neutral fraction-matter density cross bispectrum terms. The sign of the matter bispectrum is positive at all redshifts and the sign of all squeezed-limit cross bispectrum terms are also found to be positive, irrespective of the stage of reionization \cite{Majumdar2018}. Further, till the mid-stage of the reionization, the neutral fraction bispectrum depends solely on the size and number density of the ionized region, with its sign being negative (see \cite{Majumdar2018} for more details). As ionized regions form and grow with the progress of reionization, the neutral fraction bispectrum starts contributing to the 21-cm bispectrum, resulting in a reduction in its magnitude.\par
Later, during the intermediate stages of the reionization, the period after the first sign change in the bispectrum, more sources of ionizing radiation emerge and the ionized bubbles around them also grow in size. Eventually, isolated bubbles begin to overlap and connect together to form larger ionized regions. This leads to an increase in the fluctuations in the 21-cm signal and the signal becomes highly non-Gaussian, which results in the higher magnitude of the 21-cm bispectrum. However, around $\bar x_{\rm HI}=0.90$, we notice a reduction in the magnitude of the 21-cm bispectrum similar to that of the 21-cm power spectrum. This happens due to the contribution from the neutral fraction-matter density cross bispectrum terms and is further enhanced by the redshift space distortions \cite{Majumdar_2020, Gill_2023}.\par
Once half of the volume is ionized the squeezed-limit $k$-triangles bispectrum is no longer driven by the distribution and sizes of the ionized regions, rather they are controlled by the sizes and distribution of the neutral regions in an ionized background. This results in the reduction of the magnitude of the 21-cm bispectrum after it reaches a maximum around $\bar x_{\rm HI}\approx0.55$. Followed by the reduction in its amplitude, the bispectrum sign changes from positive to negative and its magnitude increases rapidly. This is because the distribution of the neutral regions in a mostly ionized IGM at the late stages of the reionization makes the signal highly non-Gaussian. As we go from $k_{1}=0.16~\text{Mpc}^{-1}$ to $k_{1}=0.33~\text{Mpc}^{-1}$ the 21-cm bispectrum becomes more and more sensitive to smaller ionized regions, thus bispectrum captures the non-Gaussianity due to the distribution of these regions resulting in higher magnitude of the 21-cm bispectrum during the intermediate stages of the reionization.
\par
We observe a decrease in the magnitude of the 21-cm bispectrum and a sign change from positive to negative during the early stages of the reionization for equilateral, stretched, L-isosceles and linear $k$-triangles, which is similar to the nature of 21-cm bispectrum for the squeezed-limit $k$-triangles. This trend in the 21-cm bispectrum of this unique $k$-triangle configuration can be explained by the same reasons we pointed out when we give our interpretation of the nature of the 21-cm bispectrum for the squeezed-limit $k$-triangles.
\subsection{Impact of reionization scenarios on specific $k$-triangles of the 21-cm bispectrum} 
\label{sec:bispec_reion_scenario}
Next, we choose a few specific $k$-triangles shapes e.g. squeezed-limit, equilateral, stretched, L-isosceles, and linear $k$-triangles, and perform a detailed comparison of the bispectrum for different reionization scenarios with respect to that of the fiducial scenario. Our choice of $k$-triangles shapes and size is set by the detectability of the bispectrum as well as the features in the bispectrum for those $k$-triangles. For the interested reader, we have presented a short discussion on the bispectrum for all unique $k$-triangle shapes and sizes and for all reionization scenarios in Appendix \ref{sec:app1}. This analysis will help us to interpret how the morphology of the IGM 21-cm signal for the different reionization scenarios impacts the 21-cm bispectrum.
\par 
First, we focus on the bispectrum for squeezed-limit $k$-triangles for different reionization scenarios for $k_{1}=0.16~\text{Mpc}^{-1}$. The left top panel of Figure~\ref{SQ_BS} shows the evolution of the 21-cm bispectrum for the squeezed-limit for $k_{1}=0.16~\text{Mpc}^{-1}$. Solid, dotted, and dashed lines are the 21-cm bispectrum for reionization scenarios which show inside-out, outside-in, and a combination of inside-out and outside-in morphologies respectively. The 21-cm bispectrum for all the reionization scenarios except the outside-in scenarios show similar features. However, these features shift in neutral fraction (i.e. redshift), and the magnitude of the 21-cm bispectrum also differs significantly from that of the fiducial scenario at different stages of the reionization.
\par
The 21-cm bispectrum for other inside-out reionization scenarios, such as clumping and PL scenarios, also shows a sign change in the bispectrum at an early stage of the reionization.
The first sign change in the bispectrum of the clumping scenario occurs around the same global neutral fraction at which the 21-cm bispectrum of the fiducial scenario also shows the sign change. However, the same sign change in PL scenarios appears at a higher neutral fraction than that of the fiducial scenario.
In the PL scenarios, large ionized bubbles are formed around the high-mass sources (i.e. massive dark matter halos) even at a very early stage of the reionization. This results in the dominant contribution by the neutral fraction bispectrum term  (which is negative) than the matter bispectrum term in the 21-cm bispectrum, which results in a rapid decrease in the magnitude of the 21-cm bispectrum and an early sign change in the bispectrum. 
\par
Later on, during the intermediate stages of reionization,  the stage at which ionized bubbles grow in size and merge to form larger ionized bubbles, the magnitude of the 21-cm bispectrum increases. During this period, the magnitude of the bispectrum is higher for PL (n=3) and PL (n=2) scenarios than the other two inside-out reionization scenarios. 
This feature can be accounted for by the presence of larger ionized regions in the PL scenarios compared to the other two inside-out reionization scenarios (see Figure~\ref{21-cm_map} for a visual inspection). The larger ionized bubbles in PL scenarios introduce stronger large-scale fluctuations in the 21-cm signal, which reflect as an enhancement of the magnitude of the large-scale 21-cm bispectrum. PL (n=3) and clumping scenarios show an early second sign change in the 21-cm bispectrum, which turns from negative to positive. In these scenarios, more than half of the volume of the IGM is fully ionized at a slightly higher neutral fraction compared to the fiducial scenario. This leads to the situation that the ionized region becomes the background, and in this background, there are neutral regions with 21-cm signal in emission, which results in the change in the sign of the bispectrum. At the late stages of the reionization, during which the distribution of the neutral regions drives the 21-cm fluctuations at large scales, the magnitude of the 21-cm bispectrum of all the reionization scenarios increases as the 21-cm signal becomes highly non-Gaussian due to the distribution of small HI regions in an ionized background.
\par
Next, we compare the 21-cm bispectrum for reionization scenarios which show a combination of inside-out and outside-in reionization morphology. At very early stages of reionization, the magnitude and sign of the 21-cm bispectrum for SXR and UV+SXR+UIB scenarios follow the signal bispectrum of the fiducial scenario and its first sign change can be explained by the toy model described to predict the sign of the signal bispectrum of the fiducial scenario. During the intermediate stages of reionization, the typical size of ionized bubbles formed in SXR and UV+SXR+UIB scenarios is smaller compared to the median bubble size of the fiducial scenario. This results in an overall reduction in the fluctuations in the 21-cm signal in these two cases. Additionally, the presence of a significant volume of partially ionized regions also reduces the strength of the 21-cm fluctuations. Thus, the magnitude of the 21-cm bispectrum for these scenarios (SXR and UV+SXR+UIB) is lower than that of the signal bispectrum for the fiducial scenario during the intermediate stages of reionization. 
\par
The second sign change in the 21-cm bispectrum for SXR and UV+SXR+UIB reionization scenarios happens at a later stage of reionization compared to that of the fiducial scenario. Note that in Section \ref{subsection:interpreting_21-cm_bispecctrum_fiducial}, we interpreted the sign change in the 21-cm bispectrum at the later stages of reionization as an indication of the fact that the 21-cm bispectrum is no longer driven by the size and distribution of the ionized regions but rather by the distribution of the neutral regions. We find that in the case of SXR and UV+SXR+UIB scenarios, more than half of the volume is fully ionized at a significantly lower global neutral fraction than that of the fiducial scenario. This implies that the 21-cm bispectrum is driven by the signal emitted from the neutral regions only at a later stage of reionization, which results in a shift in the second sign change of the bispectrum.
\par    
The evolution of the 21-cm  bispectrum for the UIB scenario, which has a complete outside-in reionization morphology, differs considerably compared to the bispectrum for the fiducial scenario. In the UIB scenario, the typical size of the ionized bubbles formed is very small compared to that of the fiducial scenario. Additionally, there are large volumes of partially ionized regions in the IGM. Due to this, the contribution from the neutral fraction field is subdominant to the large-scale 21-cm bispectrum, which results in a gradual decrease in bispectrum magnitude and delayed first sign change compared to that of the fiducial scenario. The magnitude of the 21-cm bispectrum remains significantly low even after the sign change in the signal bispectrum from positive to negative. It is also important to note that the squeezed-limit bispectrum for the UIB scenario does not show a second sign change. We have demonstrated in Section \ref{subsection:interpreting_21-cm_bispecctrum_fiducial} that the large scale 21-cm bispectrum for the squeezed limit $k$-triangle captures the relative contrast between the 21-cm signal fluctuations and the background. In the fiducial scenario, the second sign change in the 21-cm bispectrum for the squeezed-limit $k$-triangles is observed when more than half of the volume of the IGM is fully ionized. However, in the case of the UIB scenario, even at the late stages of reionization less than half of the volume of the IGM is fully ionized. Additionally, the large volume of the partially ionized regions emits the 21-cm signal with a reduced magnitude compared to the scenario if they were fully neutral. Thus, the regions with a 21-cm signal in emission still remain as the background. Due to this nature of the 21-cm morphology, the sign of the signal bispectrum in the UIB scenario remains negative.
 \par
 The timing of the second sign change in the large-scale 21-cm bispectrum for the squeezed-limit $k$-triangles can be used as an independent method to pinpoint the time at which about half of the volume of the IGM is fully ionized. In the power spectrum, this feature appears as a peak in magnitude. However, the change in the sign of the bispectrum would be more distinguishable than the peak in the power spectrum.
\par
Next, we compare the nature of the 21-cm bispectrum for equilateral $k$-triangles for $k_{1}=0.16~\text{Mpc}^{-1}$ for all of the reionization scenarios with the fiducial scenario (middle panel of Figure~\ref{SQ_BS}). At the earlier stages of the reionization, the magnitude of the 21-cm bispectrum decreases, and the sign of the bispectrum changes from positive to negative for reionization scenarios showing inside-out and combination of inside-out and outside-in morphologies. Among the reionization scenarios considered here, the earliest ``first'' sign change in signal bispectrum is observed in the PL scenarios. However, in the case of the UIB scenario, there is only a gradual decrease in the magnitude of the signal bispectrum, and the sign change appears at a significantly later stage of the reionization compared to that of the other reionization scenarios considered here. 
\par
Once the bispectrum has gone through its first sign change, its magnitude increases and reaches a maximum value. The neutral fraction at which the signal bispectrum reaches its maximum varies depending on the reionization scenario. We found that the magnitude of the bispectrum is highest for the PL scenarios during the intermediate stages of reionization. Similar to the fiducial scenario, the equilateral bispectrum does not show any second sign change during the later stages of reionization for the rest of the reionization scenarios as well.
We observe a significant change in the magnitude of the 21-cm bispectrum and the timing of the bispectrum sign change in PL scenarios with respect to that of the fiducial scenario. This difference in bispectrum behaviour is due to the presence of larger ionized regions in the PL scenarios compared to that of the fiducial scenario at any stage of reionization. As we vary $k_{1}$ from $0.16~\text{Mpc}^{-1}$ to $0.33~\text{Mpc}^{-1}$, the magnitude of the bispectrum for equilateral $k$-triangles increases during the intermediate stages of the reionization, similar to what is observed for the squeezed-limit $k$-triangles.
\par
Next, we discuss the 21-cm bispectrum for stretched $k$-triangles for all reionization scenarios (bottom panel of Figure~\ref{SQ_BS}). Here we limit our interpretation only to bispectrum for large length scales i.e. $k_1=0.16~\text{Mpc}^{-1}$. We observe that for all reionization scenarios, the magnitude of the signal bispectrum decreases as reionization progresses and eventually shows a sign change. This general feature of the bispectrum can be attributed to the gradual increase in the volume of the ionized regions. We observe that the sign change in the bispectrum for the PL scenario happens at a relatively earlier stage of the reionization compared to all other scenarios. This characteristic is the same as that of the signal bispectrum for the squeezed-limit and equilateral $k$-triangles discussed earlier and can be interpreted in the same manner. 
Out of all reionization scenarios, PL and the UIB scenarios show significantly different trends compared to that of the fiducial scenario. Beyond the first sign change, the PL (n=3) scenario shows a rapid increase in the magnitude of the bispectrum. The bispectrum reaches its maximum magnitude around $\bar x_{\rm HI}\approx0.85$, beyond which its magnitude starts to decrease. The bispectrum goes through its second sign change also at a significantly early stage of the reionization ($\bar x_{\rm HI}\approx0.72$). 
These two transition points in the case of the PL (n=3) scenario appear at a significantly higher neutral fraction than that of the fiducial scenario. A similar feature can be observed for the PL (n=2) scenario as well only shifted slightly towards lower neutral fractions compared to that of the  PL (n=3) scenario. We also observed a strong correlation between the neutral fraction at which the 21-cm bispectrum goes through the second sign change and the peak of the 21-cm power spectrum. 
\par
Finally, we discuss the 21-cm bispectrum for L-isosceles and linear $k$-triangles at large scales i.e. $k_{1}=0.16~\text{Mpc}^{-1}$ (Figure~\ref{L-isosceles_and_linear }). The general features of 21-cm bispectrum evolution remain the same in this case as that of the L-isosceles and linear $k$-triangles. Here, we also observe that the same features appear to be shifted in neutral fraction (i.e. redshift) depending on the reionization scenario under consideration, excluding the UIB scenario. Additionally, there is a difference in the magnitude of the bispectrum which varies depending on the reionization scenario under consideration. The 21-cm bispectrum for the UIB scenario shows a significant difference in sign and magnitude. The bispectrum for L-isosceles and linear $k$-triangles for the UIB scenario does not show a sign change in the bispectrum at the late stages of the reionization.
\par
In this article, we have demonstrated that the 21-cm bispectrum can distinguish between different reionization scenarios. However, it is also essential to verify whether these conclusions remain valid even in the presence of various systematic uncertainties associated with a realistic radio interferometric observation. To address this issue, we did an order-of-magnitude estimate of the uncertainty in the 21-cm bispectrum arising solely due to instrumental noise and sample variance considering observations with an instrument similar to SKA1-Low. We followed the formalism of~\cite{Soccimarro_2004, Soccimarro_1998, Murmu_2023} to estimate the variance in the 21-cm bispectrum ($\sigma^2_{N}(B)$) due to instrumental noise. The noise variance in the 21-cm bispectrum is expressed as
\begin{align}
    \sigma^2_{\rm N}(B) \approx s_B \frac{V_{f}}{V_{B}} P_{\rm N}(k_1,z) P_{\rm N}(k_2,z) P_{\rm N}(k_3,z),
\end{align}
where $V_B\approx 8 \pi^2 k_1 k_2 k_3 \Delta k_1 \Delta k_2 \Delta k_3$, $V_f = 8\pi^3 / {V_s}$ is the volume of the fundamental cell and $V_s$ is the survey volume. The variable $s_B$ takes the value $s_B = 1,2,6$ for general, isosceles and equilateral $k$-triangles, respectively. $P_{\rm N}(k,z)$ is the noise power spectrum solely due to the instrumental noise. To estimate the noise power spectrum $P_{\rm N}(k,z)$, we followed the formalism of \cite{Bull_2015}. More details on the estimation of the noise power spectrum and the choice of the instrumental parameters can be found in section (5.4) of \cite{Murmu_2023}. 
\par
In Figure~\ref{SQ_BS}, the shaded region around the line plot shows the $3\sigma$ error in the bispectrum due to instrumental noise and sample variance for 1000 hours of observation with SKA1-Low. 
We found that for $k_1$=0.16 Mpc$^{-1}$, the noise contribution is smallest at lower neutral fractions (lower redshifts) and increases for higher neutral fractions (higher redshifts). Additionally,  as we go from $k_1$=0.16 Mpc$^{-1}$ to 0.33 Mpc$^{-1}$, the noise increases for squeezed-limit, equilateral and stretched $k$-triangles. Even in the presence of noise in the bispectrum, reionization scenario-dependent features in the 21-cm bispectrum $k$-triangles are distinguishable. Here, we do not include the uncertainty contributed by the cosmic variance in the bispectrum, which will dominate on large scales. The uncertainty due to the cosmic variance in 21-cm  bispectrum estimation was studied thoroughly by~\cite{Rajeshbispectrumdetectability} for a range of length scales considering 1024 hours of observation with SKA1-Low. They concluded that the error due to cosmic variance dominates the uncertainty in the 21-cm bispectrum estimates for the scales considered in our work. However, they noted that the 21-cm bispectrum for the squeezed-limit $k$-triangles has the smallest error due to cosmic variance at these scales and has a Signal-to-Noise Ratio (SNR) higher than $5\sigma$. They also noted that the stretched and equilateral $k$-triangles would have more than $3\sigma $ detectability even when cosmic variance is considered. So we expect the model-dependent features in the 21-cm bispectrum will remain distinguishable at least for squeezed-limit $k$-triangles with 1000 hours of SKA1-Low observations. To get a more accurate estimate of the uncertainty in the 21-cm bispectrum measurements, one has to do a thorough numerical analysis, including the effect of instrumental noise and cosmic variance, as presented by~\cite{Rajeshbispectrumdetectability}. This is beyond the scope of this article and we plan to pursue this in the future.

\subsection{Quantifying the difference in 21-cm bispectrum between different reionization scenarios}
\label{sec:bispec_difference}
For a better understanding of how the magnitude and sign of the 21-cm bispectrum for various reionization scenarios are different from the fiducial case, we estimate the fractional change in the magnitude of the 21-cm bispectrum for all unique $k$-triangles for all reionization scenarios with respect to the fiducial scenario. Additionally, we also study the nature of differences in the sign of the bispectrum for various reionization scenarios.
\par
\begin{figure}
    \centering
    \includegraphics[width=\textwidth]{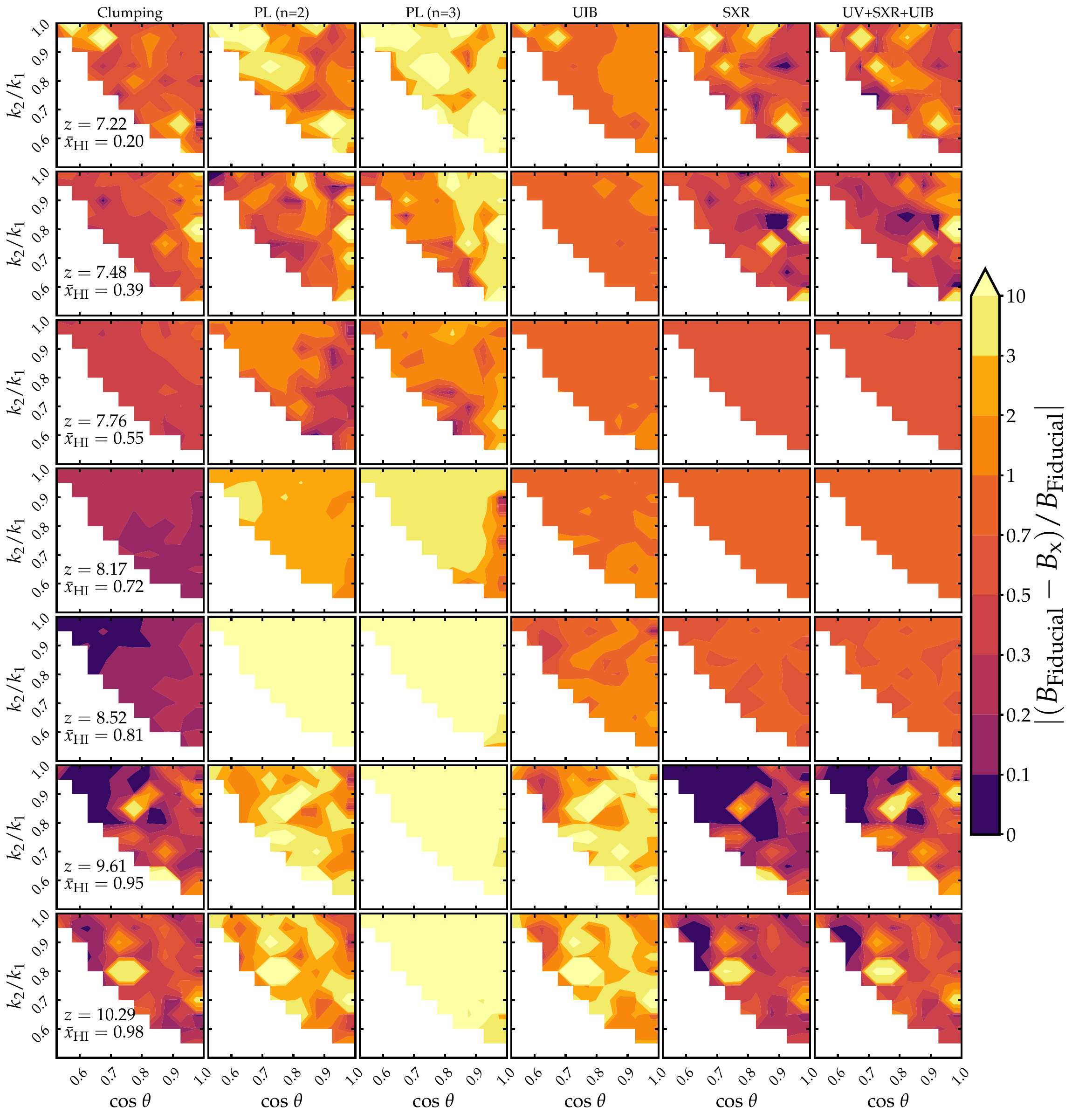}
    \caption{Relative difference in the 21-cm bispectrum between the fiducial and other reionization scenarios at all the unique $k$-triangles ($k_{1}=0.16~\text{Mpc}^{-1}$) is shown in this figure.}
    \label{bispectrum_relative_diff_plot}
\end{figure}

First, we discuss the impact of different reionization scenarios on the magnitude of the 21-cm bispectrum for all the unique $k$-triangles followed by the discussion of how they impact the sign of the 21-cm bispectrum. For this comparison, we always use the 21-cm bispectrum for the fiducial scenario as our benchmark unless mentioned otherwise. Figure~\ref{bispectrum_relative_diff_plot} shows the relative fractional difference between the magnitude of the 21-cm bispectrum for all the unique $k$-triangles for $k_{1}=0.16~\text{Mpc}^{-1}$ for the fiducial scenario and other reionization scenarios, i.e $|(B_{\text{Fiducial}}-B_{\text{X}})/B_{\text{Fiducial}}|$, where $\text{X}$ is the specific reionization scenario that is being compared. An initial inspection of Figure~\ref{bispectrum_relative_diff_plot} reveals a significant impact on the magnitude of the 21-cm bispectrum due to change in the reionization scenario. For almost all the unique $k$-triangles and stages of the reionization, the change in the magnitude of the 21-cm bispectrum is higher than 10\%, and for some of the scenarios, it is even higher than 1000\% at certain stages of the reionization.
\par
In the case of the clumping scenario, the fractional change in the 21-cm bispectrum increases during the early stages of reionization and then decreases in the intermediate stages. At later stages, this fractional change in the bispectrum keeps on increasing as the neutral fraction decreases. The change in the magnitude of the bispectrum is highest at lower neutral fractions. Among all the reionization scenarios, PL scenarios show the highest change in the magnitude of the bispectrum at all stages of the reionization. The change in the magnitude of the 21-cm bispectrum corresponding to the PL (n=3) scenario for all the unique $k$-triangles is higher than 1000\% during the early stages of the reionization. This observation can be accounted by the presence of very large ionized regions formed at the early stages of the reionization in this specific case.
\par
The reionization scenarios in between the two extreme scenarios ``inside-out'' (fiducial) and ``outside-in'' (UIB), such as SXR and UV+SXR+UIB show almost similar trends in the relative fractional change in the magnitude of bispectrum for all unique $k$-triangles and this change is highest during the intermediate stages of the reionization. In the case of the UIB scenario, which shows a completely outside-in reionization morphology, the change in the magnitude of the 21-cm bispectrum is highest at higher neutral fraction values than at lower neutral fraction values. At the early stages of the reionization, some of the $k$-triangles show a change in magnitude of the bispectrum higher than 1000\%. At all other stages of the reionization, the change in the bispectrum for most of the unique $k$-triangles are between 100\% and 300\%.
\begin{figure}
    \centering
    \includegraphics[width=\textwidth]{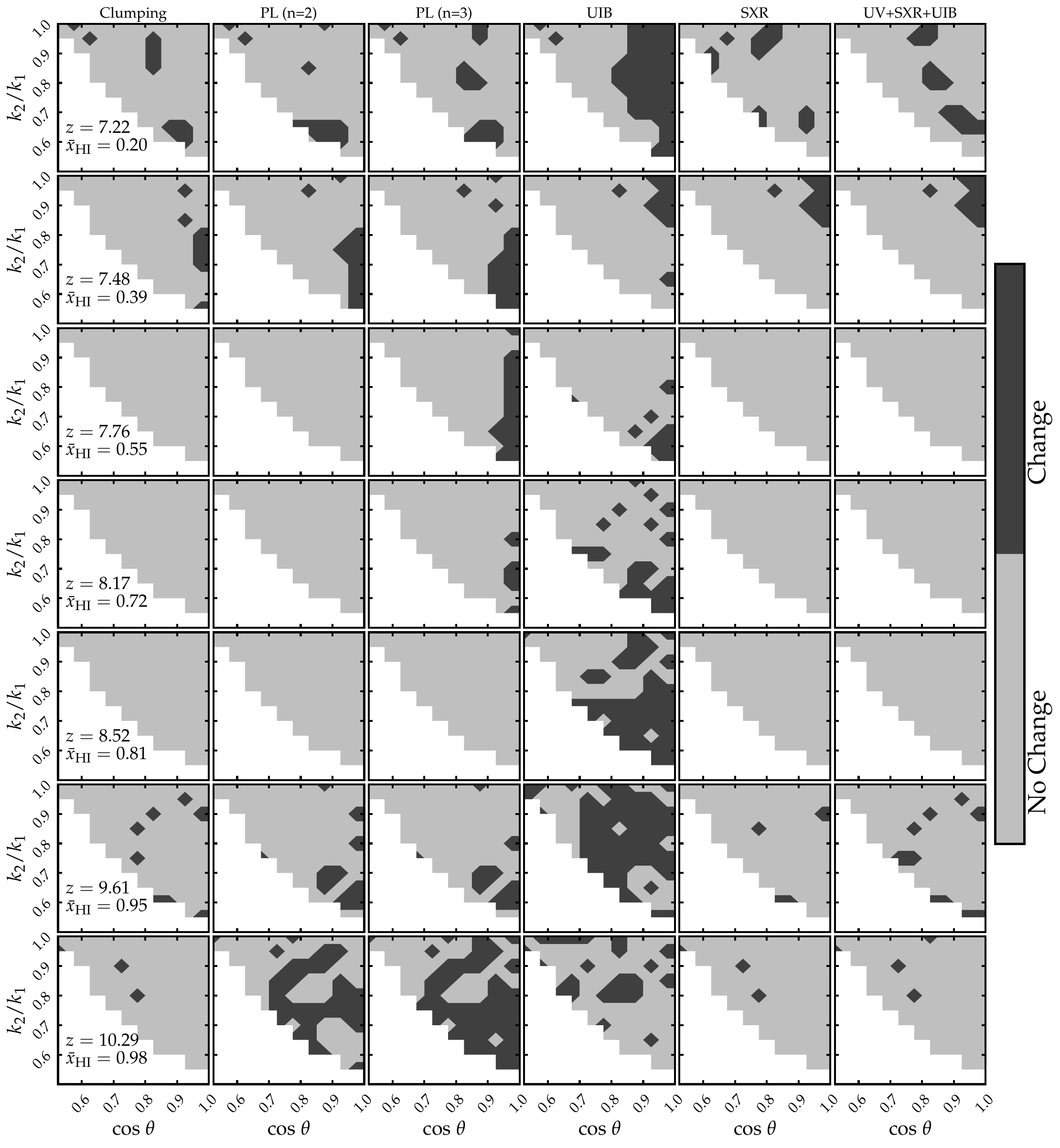}
    \caption{The change in sign of the 21-cm bispectrum between the fiducial and other reionization scenarios for all the unique $k$-triangles ($k_1=0.16~\text{Mpc}^{-1}$) is shown in this figure. ``Change'' denotes that the sign of the 21-cm bispectrum of the given reionization scenario is different from the fiducial one, whereas ``No change'' denotes that there is no change in the sign of the bispectrum.}
    \label{bispectrum_relative_diff_sign_plot}
\end{figure}
\par
The sign is also an important feature of the EoR 21-cm bispectrum and in principle can be used to distinguish between different reionization scenarios. Figure~\ref{bispectrum_relative_diff_sign_plot} shows the difference in the sign of the 21-cm bispectrum for all the reionization scenarios and all triangle shapes. The areas in black in the $t-\cos\theta$ plane, which is labelled as ``Change'' indicates that there is a change in the sign of the 21-cm bispectrum for the specific reionization scenario with respect to the fiducial scenario and the areas in grey labelled as ``No change'' indicates that there is no sign change with respect to the fiducial scenario.
\par
The area in the $t-\cos\theta$ plane with change in the bispectrum sign with respect to the fiducial scenario is smallest for clumping, SXR, and UV+SXR+UIB scenarios. However, for PL scenarios a significant area of the $t-\cos\theta$ plane of the unique $k$-triangles shows a sign change in the bispectrum with respect to the fiducial scenario at the early stages of the reionization. This is because in PL scenarios the contribution from the neutral fraction bispectrum to the 21-cm bispectrum starts to dominate already at an early stage of the reionization due to the early formation of large ionized regions. This results in an early sign change from positive to negative in the 21-cm bispectrum for PL scenarios. Unlike other reionization scenarios, in the case of the UIB scenario, a significant area of the $t-\cos\theta$ plane shows a sign change at almost all stages of reionization and the area in the $t-\cos\theta$ plane with sign a change is largest at the early and the late stages of reionization.
\par
The results demonstrated here provide us an avenue to address the question that we posed at the outset i.e. is the additional information content in the 21-cm signal bispectrum significant enough to outperform the power spectrum in distinguishing different reionization scenarios? The power spectrum, by definition, cannot quantify the non-Gaussianity present in the signal fluctuations. The change in the reionization scenario results only in a change in the shape of the power spectrum. Additionally, the power spectrum by definition is always positive. On the contrary, the bispectrum can quantify the time-evolving non-Gaussianity present in the signal fluctuations and it can take both negative and positive values. Along with the change in the magnitude of the 21-cm bispectrum (with redshift, scale, and triangle shape), the sequence of sign changes in the bispectrum and the timing of sign changes enables one to distinguish these reionization scenarios better than the 21-cm power spectrum can.
\section{Summary and discussions}
\label{section: summary}
The morphology of the IGM 21-cm signal during the EoR will vary depending on the properties of the sources of ionizing radiation and the physical processes within the IGM. We expect the differences in 21-cm morphologies arising from different reionization scenarios will have a significant impact on the 21-cm signal fluctuations, which in turn will affect the features in the signal bispectrum that have not been studied to date ~\cite{Majumdar2018,Hutter_2020,Majumdar_2020,Raste_2023,Gill_2023}.
In this work, we study the impact of variation in the morphology of the IGM 21-cm maps during the EoR (arising due to the change in the ionizing source properties and the physical processes within the IGM) on the 21-cm bispectrum. For this study, we considered a set of semi-numerical reionization simulations that differ by how the number of ionizing photons is related to the host halo mass and the distribution of the rest frame energy of the ionizing photons. Our primary focus is on how well the 21-cm bispectrum can distinguish between these
reionization scenarios showing inside-out, outside-in, and a combination of inside-out and outside-in morphologies. We estimated the 21-cm bispectrum for all the unique $k$-triangles for all the reionization scenarios. Focusing on a few specific $k$-triangle shapes, our analysis further shows how the sign and magnitude of the large scale (low $k_{1}$ values) 21-cm bispectrum for different reionization scenarios vary with the stage of reionization.
\par
Our study shows that the shape, sign, and magnitude of the 21-cm bispectrum can distinguish between different reionization 
scenarios. We summarise our main findings below:
\begin{itemize}
    \item The large-scale 21-cm bispectrum for squeezed-limit $k$-triangles can distinguish between reionization scenarios, which show inside-out (fiducial scenario) and outside-in (UIB scenario) morphologies. The 21-cm bispectrum for squeezed-limit $k$-triangles  for the fiducial scenario shows two sign changes. The change in the sign of the 21-cm bispectrum at an early stage of the reionization can be interpreted as a signature of the onset of the reionization. The second sign change of the 21-cm bispectrum, observed at the late stages of the reionization when more than half of the volume of the IGM is ionized, indicates the presence of the neutral regions with 21-cm signal in emission in an ionized background. However, the 21-cm bispectrum for squeezed-limit $k$-triangles for the outside-in reionization scenario (which is the UIB scenario) does not show a change in the sign of the bispectrum at the late stages of the reionization. Additionally, the bispectrum for squeezed-limit $k$-triangles of the outside-in reionization scenario shows a delayed first sign change compared to the bispectrum of the fiducial scenario (especially for $k_{1} = 0.16~\text{Mpc}^{-1}$) and after the sign change the magnitude of the bispectrum is lower compared to the fiducial scenario. These features observed in the 21-cm bispectrum can be used to distinguish between inside-out and outside-in reionization morphologies. Similar features are present in the linear $k$-triangles and the vicinity of the squeezed-limit $k$-triangles.
    \item Our analysis shows that the magnitude and sign change of the large-scale 21-cm bispectrum can distinguish between different inside-out reionization scenarios and also reionization scenarios which are a combination of inside-out and outside-in morphology. In comparison with the fiducial scenario, the 21-cm bispectrum for the squeezed-limit $k$-triangles for PL scenarios (inside-out scenario) shows early first and second sign changes and higher magnitude. In the case of SXR and UV+SXR+UIB scenarios (combination of inside-out and outside-in), the second sign change of the squeezed-limit 21-cm bispectrum occurs at a later stage of the reionization. Additionally, the magnitude of the 21-cm bispectrum for these scenarios is lower than the fiducial case. We further observe that these features are also present in the linear $k$-triangles and in the vicinity of the squeezed-limit $k$-triangles.
    \item We did an order-of-magnitude estimate of the uncertainty in the 21-cm bispectrum due to instrumental noise for squeezed-limit, equilateral and stretched $k-\text{triangles}$. We found that these bispectra can distinguish different reionization scenarios even in the presence of instrumental noise. However, in order to get an accurate estimate of the uncertainty in the 21-cm bispectrum estimation, one has to do a thorough numerical analysis, including both instrumental noise and cosmic variance, as presented in~\cite{Rajeshbispectrumdetectability}. This is beyond the scope of the present article, as it will require a significant amount of computing resources, given the volume and mass resolution of the simulation that has been considered here.
    \item 
    The sign of the large-scale 21-cm bispectrum of the squeezed-limit $k$-triangles changes from negative to positive when more than half of the volume of the IGM is fully ionized. This appears to be robust feature across all the reionization scenarios considered in our work. The timing of this sign change in the bispectrum for the squeezed-limit $k$-triangles thus can be used as an independent way to pinpoint the time at which about half of the volume of the IGM is fully ionized.
    \item 
    We found that there is a significant change in the magnitude and sign of the 21-cm bispectrum for all the unique $k$-triangles with the change in the morphology of the IGM 21-cm signal which is quantified in Figures \ref{bispectrum_relative_diff_plot} and \ref{bispectrum_relative_diff_sign_plot}. The change in the magnitude of the 21-cm bispectrum for different reionization scenarios with respect to the fiducial scenario is found to be higher than 10\% for almost all the unique $k$-triangles for all stages of reionization. In the case of PL scenarios, the change in the magnitude of the 21-cm bispectrum is even higher than 1000\% at certain stages of the reionization. We also observe that the sign of the 21-cm bispectrum for different reionization scenarios is different from the fiducial scenario at different stages of reionization. This feature can also be used to distinguish between different reionization scenarios.
    \item The 21-cm bispectrum for the $k$-triangles corresponding to the squeezed-limit, in the vicinity of the squeezed-limit and the linear shapes show more features than other unique $k$-triangles. However, we observe a significant change in the magnitude of the 21-cm bispectrum for different reionization scenarios for all the unique $k$-triangles. This points towards the fact that we need a detailed study of the information content in the 21-cm bispectrum with varying sizes and shapes of $k$-triangle configurations to be able to identify the best unique $k$-triangle configurations for distinguishing different reionization morphologies.
    \item The 21-cm power spectrum of these different reionization scenarios also shows changes in amplitude, with the highest change being at the mid-stages of the reionization. However, the power spectrum, by definition, cannot quantify the non-Gaussianity present in the signal fluctuations. Additionally, the sign of the signal power spectrum is always positive. The bispectrum can quantify the time-evolving non-Gaussianity present in the signal fluctuations and the bispectrum can take both negative and positive values. Along with the change in the magnitude of the 21-cm bispectrum (with redshift, scale, and triangle shape), the sequence of sign changes in the bispectrum and the timing of sign changes enables one to distinguish these reionization scenarios better than the 21-cm power spectrum.
\end{itemize}
In this work, we did not include the effect of line-of-sight anisotropy arising due to the light-cone effect on the 21-cm signal~\cite{Datta_2012,Datta_2014}, which have been found to have a significant impact on the EoR 21-cm bispectrum~\cite{Rajeshbispectrumdetectability}. Estimating the EoR 21-cm bispectrum from the actual signal is challenging due to the presence of the residual foregrounds. In this work, we have not considered the effect of residual foregrounds on the 21-cm bispectrum. Furthermore, we also did not include a comprehensive study of the detectability of the 21-cm bispectrum for all unique $k-\text{triangles}$ for these reionization scenarios by the ongoing and upcoming interferometric observations by including the thermal noise~\cite{Yoshiura_2015}, uncertainty due to cosmic variance~\cite{Rajeshbispectrumdetectability} and other systematic uncertainties associated with the interferometric observations. We plan to address these issues in our future work.

\appendix 
\acknowledgments
LN acknowledges the financial support by the Department of Science and Technology, Government of India, through the INSPIRE Fellowship. SM and LN acknowledge financial support through the project titled ``Observing the Cosmic Dawn in Multicolour using Next Generation Telescopes'' funded by the Science and Engineering Research Board (SERB), Department of Science and Technology, Government of India through the Core Research Grant No. CRG/2021/004025. SM and LN acknowledge the use of computing infrastructure for this work, which is hosted at the DAASE, IIT Indore and was procured through the funding via , Department of Science and Technology, Government of India sponsored DST-FIST grant no. SR/FST/PSII/2021/162 (C) awarded to the DAASE, IIT Indore. 
SM and II acknowledge financial support through the project titled ``Galaxies through the ages: using synergistic multi-wavelength observations'' (Project No. P2505) funded under the ``Scheme for Promotion of Academic and Research Collaboration (SPARC)'' from the Ministry of Education, India. CSM acknowledges funding from the Council of Scientific and Industrial Research (CSIR) under the grant 09/1022(0080)/2019-EMR-I. RG acknowledges support from the Kaufman Foundation (Gift no. GF01364). GM's research has been supported by Swedish Research Council grant 2020-04691$\_\text{VR}$.
\appendix
\section{Bispectrum for all the unique $k$-triangles}
\label{sec:app1}
\begin{figure}
    \centering
    \includegraphics[width=\textwidth]{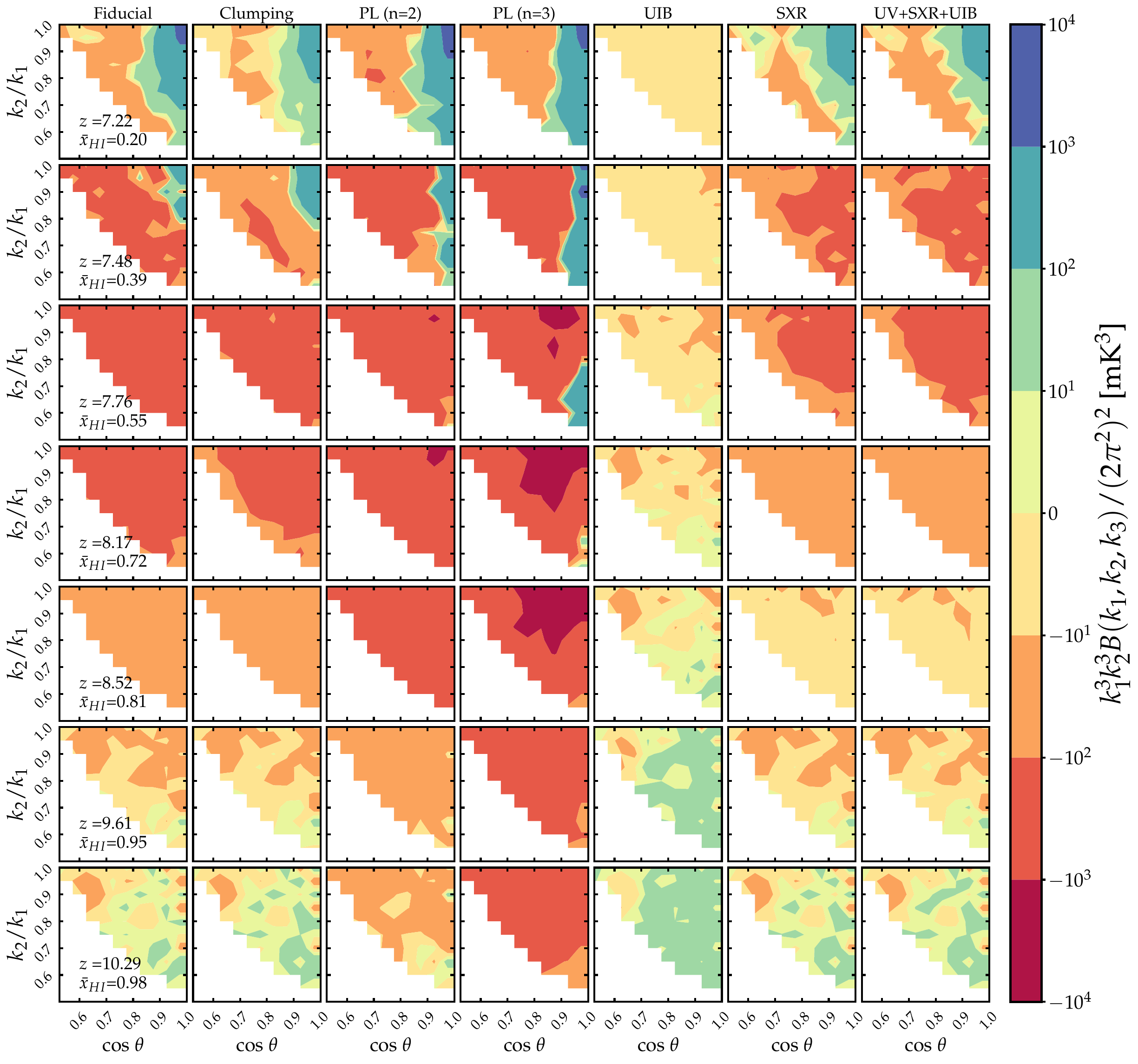}
    \caption{Evolution of the 21-cm bispectrum for all unique k-triangles for different reionization scenarios with global neutral fraction for $k_{1}=0.16~\text{Mpc}^{-1}$ is shown in this figure.}
    \label{unique_triangles}
\end{figure}

Figure~\ref{unique_triangles} shows the 21-cm bispectrum for all reionization scenarios and unique $k$-triangle shapes for $k_{1}=0.16~\text{Mpc}^{-1}$. First, we show the nature of the 21-cm bispectrum for the fiducial scenario. At the early stages of the reionization, the bispectrum in most of the area of the $t-\cos\theta$ plane is positive. As reionization progresses, the bispectrum in this entire plane becomes negative, and the magnitude of the bispectrum increases until the mid-stage of the reionization is reached. During the late stages of the reionization, we observe a reduction in the magnitude of the bispectrum. We observe a change in the sign of the bispectrum for some specific $k$-triangles shapes, such as the squeezed-limit, the triangles in the vicinity of the squeezed-limit, and the linear $k$-triangles. The 21-cm bispectrum for all unique $k$-triangles and all of the reionization scenarios except the UIB scenario show almost similar features to that of the fiducial case. However, these features shift in neutral fraction (i.e. redshift), and we also observe a difference in the magnitude of the bispectrum. 


\bibliographystyle{JHEP}
\bibliography{biblio}
\end{document}